\documentclass[twocolumn,showpacs,superscriptaddress,nofootinbib]{revtex4-1}

\usepackage[latin9]{inputenc}
\usepackage{amsmath}
\usepackage{amssymb}
\usepackage{amsthm}
\usepackage{graphicx}
\usepackage{color}

\theoremstyle{plain}
\newtheorem*{theorem*}{Theorem}

\makeatletter

\@ifundefined{textcolor}{}
{
 \definecolor{BLACK}{gray}{0}
 \definecolor{WHITE}{gray}{1}
 \definecolor{RED}{rgb}{1,0,0}
 \definecolor{GREEN}{rgb}{0,1,0}
 \definecolor{BLUE}{rgb}{0,0,1}
 \definecolor{CYAN}{cmyk}{1,0,0,0}
 \definecolor{MAGENTA}{cmyk}{0,1,0,0}
 \definecolor{YELLOW}{cmyk}{0,0,1,0}
}

\usepackage{color}
\usepackage{amsfonts}
\usepackage{graphics}
\usepackage{bm} 
\usepackage{epsfig}\usepackage{amsthm}

\usepackage{lipsum}

\newcommand\blfootnote[1]{%
  \begingroup
  \renewcommand\thefootnote{}\footnote{#1}%
  \addtocounter{footnote}{-1}%
  \endgroup
}

\def\identity{\leavevmode\hbox{\small1\kern-3.8pt\normalsize1}}

\renewcommand{\epsilon}{\varepsilon}

\makeatother

\begin{document}

\title{Probing quantum features of photosynthetic organisms}

\author{Tanjung Krisnanda}
\affiliation{School of Physical and Mathematical Sciences, Nanyang Technological University, 637371 Singapore, Singapore}
\blfootnote{Correspondence: T. Krisnanda (tanjungkrisnanda@gmail.com) or T. Paterek (tomasz.paterek@ntu.edu.sg)}

\author{Chiara Marletto}
\affiliation{Clarendon Laboratory, University of Oxford, Parks Road, Oxford OX1 3PU, United Kingdom}

\author{Vlatko Vedral}
\affiliation{Clarendon Laboratory, University of Oxford, Parks Road, Oxford OX1 3PU, United Kingdom}
\affiliation{Centre for Quantum Technologies, National University of Singapore, 117543 Singapore, Singapore}

\author{Mauro Paternostro}
\affiliation{School of Mathematics and Physics, Queen's University, Belfast BT7 1NN, United Kingdom}

\author{Tomasz Paterek}
\affiliation{School of Physical and Mathematical Sciences, Nanyang Technological University, 637371 Singapore, Singapore}
\affiliation{MajuLab, CNRS-UCA-SU-NUS-NTU International Joint Research Unit, UMI 3654 Singapore, Singapore}

\begin{abstract}
Recent experiments have demonstrated strong coupling between living bacteria and light.
Here we propose a scheme capable of revealing non-classical features of the bacteria (quantum discord of light-bacteria correlations) without exact modelling of the organisms and their interactions with external world.
The scheme puts the bacteria in a role of mediators of quantum entanglement between otherwise non-interacting probing light modes.
We then propose a plausible model of this experiment, using recently achieved parameters, demonstrating the feasibility of the scheme.
Within this model we find that the steady state entanglement between the probes, which does not depend on the initial conditions, 
is accompanied by entanglement between the probes and bacteria, and provides independent evidence of the strong coupling between them.
\end{abstract}

\maketitle

\section*{Introduction} 
There is no a priori limit on the complexity, size or mass of objects to which quantum theory is applicable.
Yet, whether or not the physical configuration of macroscopic systems could showcase quantum coherences has been the subject of a long-standing debate. 
The pioneers of quantum theory, such as Schr\"odinger \cite{SCH} and Bohr \cite{BOH}, wondered whether there might be limitations to living systems obeying the laws of quantum theory. 
Wigner even claimed that their behaviour violates unitarity~\cite{WIG}. 

A striking way to counter such claims on the implausibility of macroscopic quantum coherence would be the successful preparation of quantum superposition states of living objects. 
A direct route towards such goal is provided by matter-wave interferometers, which have already been instrumental in observing quantum interference from complex molecules \cite{arndt}, and are believed to hold the potential to successfully show similar results for objects as large as viruses in the near future.

However, other possibilities exist that do not make use of interferometric approaches. 
An instance of such alternatives is to interact a living object with a quantum system in order to generate quantum correlations.
Should such correlations be as strong as entanglement, measuring the quantum system in a suitable basis could project the living object into a quantum superposition. 
Furthermore, requesting the establishment of entanglement is, in general, not necessary as the presence of quantum discord, that is a weaker form of quantum correlations, would already provide evidence that the Hilbert space spanned by the living object must contain quantum superposition states~\cite{discord,discord2,discord3,discord4,discord5}.
For example, by operating on the quantum system alone one could remotely prepare quantum coherence in the living object~\cite{coherence-distillation}.

A promising step in this direction, demonstrating strong coupling between living bacteria and optical fields and suggesting the existence of entanglement between them~\cite{bacteria-th}, has recently been realised~\cite{bacteria-exp}.
See also Refs. \cite{bio1,bio2,bio3,bio4,bio5,bio6,bio7,bio8,bio9,bio10,bio11} for a broader picture of quantum effects in photosynthetic organisms.
However, the experimental results reported in Ref.~\cite{bacteria-exp} can as well be explained by a fully classical model~\cite{ZHU, bacteria-th,bacteria-exp,rabienergy}, which calls loud for the design of a protocol with more conclusive interpretation.

In this paper we make a proposal in such a direction by designing a thought experiment in which the bacteria are mediating interactions between otherwise uncoupled light modes.
This scheme fits into the general framework of Ref.~\cite{revealing}, which shows in the present context that quantum entanglement between the light modes can only be created if the bacteria are non-classically correlated with them during the process.
It is important to realise that in this way we bypass the need of exact modelling of the living organisms and their interactions with external world.
Indeed, experimenters are never asked to directly operate on the bacteria, it is solely sufficient to observe the light modes.
A positive result of this experiment, i.e. observation of quantum entanglement between the light modes, provides an unambiguous witness of quantum correlations, in the form of quantum discord, between the light and bacteria.

In order to demonstrate that there should be observable entanglement in the experiment we then propose a plausible model of light-bacteria interactions and noises in the experiment.
We focus on the optical response of the bacteria and model their light-sensitive part by a collection of two-level atoms with transition frequencies matching observed bacterial spectrum~\cite{bacteria-exp}.
All processes responsible for keeping the organisms alive are thus effectively put into the environment of these atoms.
We argue that standard Langevin approach gives a sensible treatment of this environment due to its quasi-thermal character, low energies compared to optical transitions and no evidence for finite-size effects.
Within this model we find scenarios with non-zero steady-state entanglement between the light modes which is always accompanied by light-bacteria entanglement (in addition to quantum discord), which is in turn empowered by the strong coupling between such systems. 

\section*{Results}

\subsection{Thought experiment}

\begin{figure}[!b]
\includegraphics[width=0.43\textwidth]{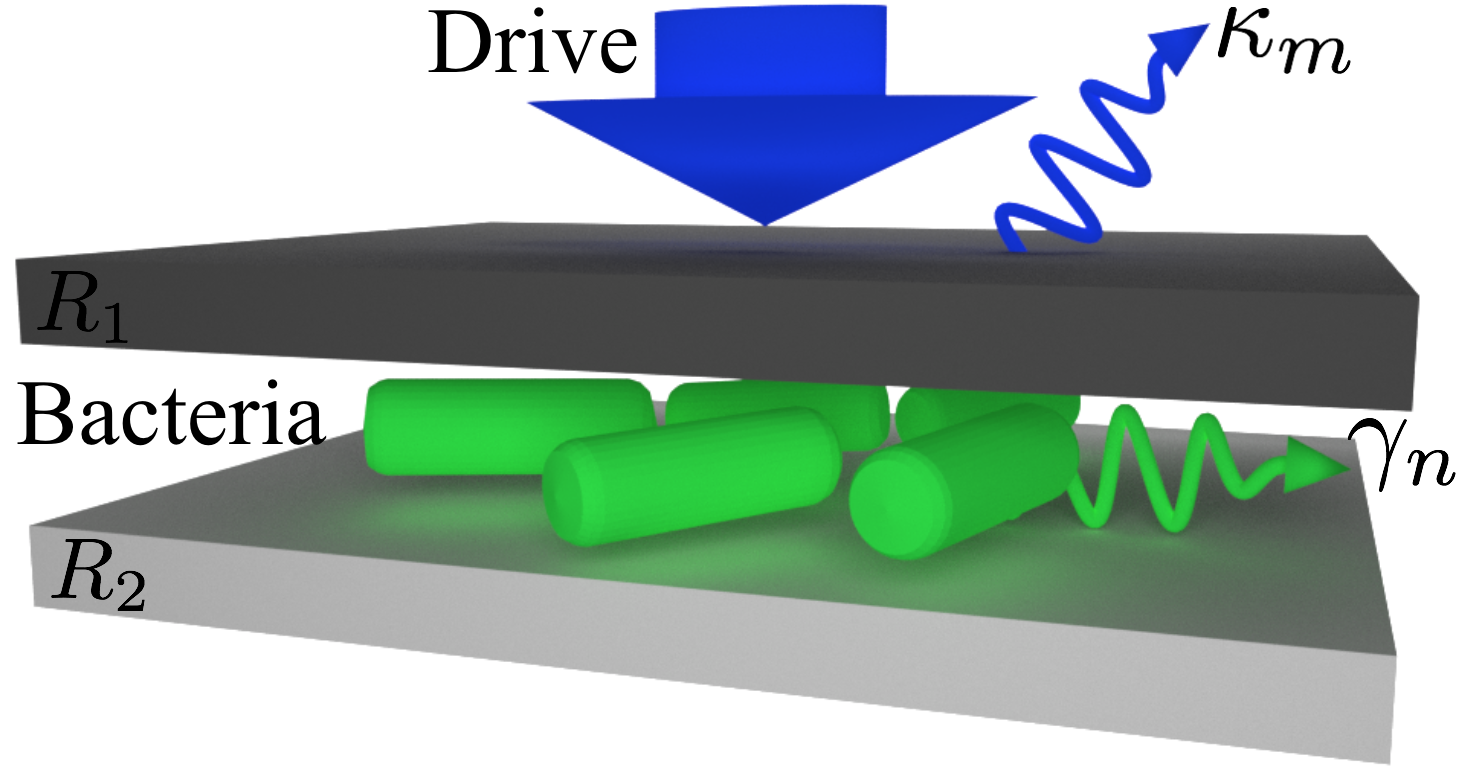}
\caption{Experimental setup revealing quantum features of photosynthetic organisms. We consider a driven single-sided multimode Fabry-Perot cavity embedding green sulphur bacteria. Here, $R_1$ is the reflectivity of the input mirror, while the end mirror is perfectly reflecting with $R_2\approx 1$.
A few cavity modes individually interact with the bacteria, but not with each other. 
Both the bacteria and cavity modes are open systems.
In particular, the interaction between the bacteria and their environment results in the energy decay rate $2\gamma_n$.
The $m^{\text{th}}$ cavity field mode experiences energy dissipation at a rate $2\kappa_m$.}
\label{FIG_ABC} 
\end{figure}

Our idea is to design a setup which, on one hand, is close to what has already been realised with bacteria and light, in order to utilise their strong coupling,
and whose description, on the other hand, can be phrased within the framework of Ref.~\cite{revealing}.
It was shown there that two physical systems, $A$ and $B$, coupled via a mediator $C$, i.e. described by a total Hamiltonian of the form $H_{AC} + H_{BC}$,
can become entangled only if quantum discord $D_{AB|C}$ is generated during the evolution.
This also holds if each system is allowed to interact with its own local environment.
Therefore, observation of quantum entanglement between $A$ and $B$ is a witness of quantum discord $D_{AB|C}$ during the evolution if one can ensure the following conditions:
\begin{itemize}
\item[(i)] $A$ and $B$ do not interact directly, i.e. there is no term $H_{AB}$ in the total Hamiltonian.
\item[(ii)] All environments are local, i.e. they do not interact with each other.
\item[(iii)] The initial state is completely unentangled (otherwise entanglement between $A$ and $B$ can grow via classical $C$~\cite{revealing}).
\end{itemize}
We now propose a concrete scheme for revealing non-classicality of the bacteria and argue how it meets these conditions.
Consider the arrangement in Fig.~\ref{FIG_ABC}.
The bacteria are inside a driven single-sided multimode Fabry-Perot cavity where they interact independently with a few cavity modes.
The cavity modes are divided into two sets which play the role of systems $A$ and $B$ in the general framework.
The bacteria are mediating the interaction between the modes and hence they represent system $C$.
Condition (i) above can be realised in practice in at least two ways.
An experimenter could utilise the polarisation of electromagnetic waves and group optical modes polarised along one direction to system $A$ and those polarised orthogonally to system $B$.
Another option, which we will study in detail via a concrete model below, is to choose different frequency modes and arbitrarily group them into systems $A$ and $B$.
Condition (ii) holds under typical experimental circumstances where the environment of the cavity modes is outside the cavity
whereas that of the bacteria is inside the cavity or even part of bacteria themselves.
The electromagnetic environment outside the cavity is a large system giving rise to the decay of cavity modes but having no back-action on them.
Therefore each cavity mode decays independently and cannot get entangled via interactions with the electromagnetic environment.
Finally, condition (iii) is satisfied right before placing the bacteria into the cavity, because at this time all three systems $A$, $B$, and $C$ are in a completely uncorrelated state $\rho_A \otimes \rho_B \otimes \rho_C$.

We note again that this discussion is generic with almost no modelling of the involved systems.
In particular, nothing has been assumed regarding the physics of the bacteria and their interactions with light and the external world.
This makes our proposal experimentally attractive.
Note also that one can think of the bacteria as a channel between the cavity modes $A$ and $B$.
The method then detects non-classicality of this channel~\cite{gyongyosi2018survey,imre2012advanced}.

In order to make concrete predictions about the amount of intermodal entanglement $E_{A:B}$ we now study a specific model for the energy of the discussed system.
This additional assumption about the overall Hamiltonian will allow us to demonstrate that the entanglement $E_{A:B}$ is accompanied by light-bacteria entanglement $E_{AB:C}$.
This independently confirms the presence of light-bacteria discord as entanglement is a stronger form of quantum correlations than discord~\cite{discord3,discord4,discord5}.
In the remainder of the paper we will therefore only calculate entanglement.

\subsection{Model}

We consider a photosynthetic bacterium, \emph{Chlorobaculum tepidum}, that is able to survive in extreme environments with almost no light~\cite{blankenship1995antenna}. Each bacterium, which is approximately $2\mu\mbox{m}\times500\mbox{nm}$ in size, contains $200-250$ chlorosomes, each having $200,000$ bacteriochlorophyll $c$ (BChl $c$) molecules. Such pigment molecules serve as excitons that can be coupled to light \cite{bacteria-exp,chlorosomes}. 
The extinction spectrum of the bacteria (BChl $c$ molecules) in water shows two pronounced peaks, at wavelengths $\lambda_{\scriptsize\mbox{I}} = 750$nm and $\lambda_{\scriptsize\mbox{II}} = 460$nm (see Fig.~1b of Ref.~\cite{bacteria-exp}).
We therefore model the light-sensitive part of the bacteria by two collections of $N$ two-level atoms with transition frequencies $(\Omega_{\scriptsize\mbox{I}},\Omega_{\scriptsize\mbox{II}})=(2.5,4.1)\times 10^{15}$ Hz.
Simplification of this model to atoms with a single transition frequency was already shown to be able to explain the results of recent experiments~\cite{bacteria-exp, chlorosomes}.
This simplification was adequate because only one cavity mode was relevant in the previous experiments.
In contrast, several cavity modes are required for the observation of intermodal entanglement and it is correspondingly more accurate to include also all relevant transitions of BChl $c$ molecules.
We assume that the molecules (two-level atoms in our model) are coupled through a dipole-like mechanism to each light mode. 
For $N\gg1$, such collections of two-level systems can be approximated to spin $N/2$ angular momenta. In the low-excitation approximation (which we will justify later), such angular momentum can be mapped into an effective harmonic oscillator through the use of the Holstein-Primakoff transformation~\cite{HP}.
This allows us to cast the energy of the overall system as
\begin{eqnarray}\label{EQ_H}
H&=&\sum_m \hbar \omega_m \hat a^{\dagger}_m \hat a_m + \sum_{n}\hbar \Omega_n \hat b^{\dagger}_n \hat b_n  \nonumber \\
&&+\sum_{m,n} \hbar G_{mn}(\hat a_m+\hat a_m^{\dagger})(\hat b_n+\hat b^{\dagger}_n) \nonumber \\ 
&&+\sum_{m} i\hbar E_m (\hat a_m^{\dagger}e^{-i\Lambda_{m} t}-\hat a_m e^{i\Lambda_{m}t}).
\end{eqnarray}
Here, $m=1,\dots,M$ is the label for the $m^\text{th}$ cavity mode, whose annihilation (creation) operator is denoted by $\hat a_m$ ($\hat a_m^\dag$) and having frequency $\omega_m$. 
Moreover, both harmonic oscillators describing the bacteria are labelled by $n=\scriptsize\mbox{I},\scriptsize\mbox{II}$ with $\hat b_n$ ($\hat b^\dag_n$) denoting the corresponding bosonic annihilation (creation) operator.
Each oscillator is coupled to the $m^\text{th}$ cavity field at a rate $G_{mn}$. 
The collective form of the coupling allows us to write $G_{mn}=g_{mn}\sqrt{N}$ with $g_{mn}=\mu_n \sqrt{\omega_m/2\hbar \epsilon_r\epsilon_0 V_m}$, where $\mu_n$ is the dipole moment of the $n^{\scriptsize \mbox{th}}$ two-level transition, $\epsilon_r$ relative permitivity of medium, and $V_m$ the $m^\text{th}$ mode volume~\cite{rabienergy}, see also \cite{optoatoms,bai2016robust} for similar treatments. 
The cavity is driven by a multimode laser, each mode having frequency $\Lambda_{m}$, amplitude $E_m=\sqrt{2P_m\kappa_m/\hbar \Lambda_{m}}$, power $P_m$, and amplitude decay rate of the corresponding cavity mode $\kappa_m$. 
It is important to notice that in Eq.~(\ref{EQ_H}) we have not invoked the rotating-wave approximation but actually retained the counter-rotating terms $\hat a_m\hat b_n$ and $\hat a_m^{\dagger}\hat b_n^{\dagger}$. These cannot be ignored in the regime of strong coupling  and we will show that they actually play a crucial role in our proposal.

We assume the local environment of the light-sensitive part of the bacteria to give rise to Markovian open-system dynamics, which is modelled as decay of the two-level systems.
For justification we note that in actual experiments the bacteria are surrounded by water which can be treated as a standard heat bath and although the environment of interest cannot be in a thermal state (because the bacteria are alive) its state is expected to be quasi-thermal.
Given that the bacterial environment of the BChl $c$ molecules is of finite size we should also justify the Markovianity assumption. 
To the best of our knowledge there is no experimental evidence against this assumption. 
Likely this is due to the fact that all excitations arriving at this environment are further rapidly dissipated to the large thermal environment of water, whose energy is small compared to the optical transitions.

We treat the environment of the cavity modes as the usual electromagnetic environment outside the cavity~\cite{noise,noise2}.
This results in independent decay rates of each mode.
Taken all together, the dynamics of the optical modes and bacteria can be written using the standard Langevin formulation in Heisenberg picture. 
This gives the following equations of motion, taking into account noise and damping terms coming from interactions with the local environments
\begin{equation}
\label{Eq_lgvn}
\begin{aligned}
\dot {\hat {a}}_m&=-(\kappa_m+i\omega_m)\hat a_m - i\sum_{n}G_{mn}(\hat b_n+\hat b_n^{\dagger})+E_m e^{-i\Lambda_{m}t}\\
&+\sqrt{2\kappa_m}\: \hat F_m,\\
\dot {\hat {b}}_n&=-(\gamma_n+i\Omega_n)\hat b_n-i\sum_{m}G_{mn}(\hat a_m+\hat a_m^{\dagger})+\sqrt{2\gamma_n}\:\hat Q_n, 
\end{aligned}
\end{equation}
where $\gamma_n$ is the amplitude decay rate of the bacterial system. 
$\hat F_m$ and $\hat Q_n$ are operators describing independent zero-mean Gaussian noise affecting the $m^\text{th}$ cavity field and the $n^\text{th}$ bacterial mode respectively.
The only nonzero correlation functions between these noises are $\langle \hat F_m(t)\hat F_{m'}^{\dagger}(t^{\prime}) \rangle=\delta_{mm'}\delta(t-t^{\prime})$ 
and $\langle \hat Q_n(t) \hat Q_{n'}^{\dagger}(t^{\prime}) \rangle=\delta_{nn'}\delta(t-t^{\prime})$~\cite{noise,noise2}.
We note that in this model the light-sensitive part of the bacteria is treated collectively, i.e. all its two-level atoms are indistinguishable.
This assumption is standardly made in present-day literature, see e.g.~\cite{bacteria-exp,chlorosomes} where modelling of the bacteria / chlorosomes as a harmonic oscillator fits observed experimental results.
But it should be stressed that this assumption deserves an in-depth experimental assessment.

We express the Langevin equations in terms of mode quadratures. 
In particular, by using $ \hat x_m\equiv ( \hat a_m+ \hat a_m^{\dagger})/\sqrt{2}$ and $ \hat y_m\equiv ( \hat a_m-\hat a_m^{\dagger})/i\sqrt{2}$ one gets a set of Langevin equations for the quadratures that can be written in a matrix equation $\dot u(t)=Ku(t)+l(t)$ with the vector $u=( \hat x_{1}, \hat y_{1},\cdots, \hat x_{M}, \hat y_{M}, \hat x_{\scriptsize\mbox{I}}, \hat y_{\scriptsize\mbox{I}},\hat x_{\scriptsize\mbox{II}}, \hat y_{\scriptsize\mbox{II}})^T$. 
Here, $K$ is a square matrix with dimension $2(M+2)$ describing the drift and $l$ is a $2(M+2)$ vector containing the noise and pumping terms (see the Methods section for explicit expressions).
The solution to the Langevin equations is given by
\begin{eqnarray}\label{EQ_Lsol}
u(t)&=&W_+(t)u(0)+W_+(t)\int_0^t dt^{\prime}  W_-(t^{\prime})l(t^{\prime}),
\end{eqnarray}
where $W_{\pm}(t)=\exp{(\pm Kt)}$. 

One can construct the covariance matrix as a function of time $V(t)$ from Eq. (\ref{EQ_Lsol}) (cf. Methods section).
Time evolution of important quantities can then be calculated from the covariance matrix, e.g. entanglement and excitation number (cf. Methods section).
We shall only be interested in the steady state, which is guaranteed when all real parts of the eigenvalues of $K$ are negative.
In this case the covariance matrix satisfies Lyapunov-like equation
\begin{equation}\label{EQ_Css}
K\:V(\infty)+V(\infty)\:K^T+D=0,
\end{equation}
where $D=\mbox{Diag}[\kappa_1,\kappa_1,\cdots,\kappa_M,\kappa_M,\gamma_{\scriptsize\mbox{I}},\gamma_{\scriptsize\mbox{I}},\gamma_{\scriptsize\mbox{II}},\gamma_{\scriptsize\mbox{II}}]$.
Note that the steady-state covariance matrix does not depend on the initial conditions, i.e. $V(0)$. 
Moreover, as the Langevin equations are linear and due to the gaussian nature of the quantum noises, the dynamics of the system is preserving gaussianity. Therefore the steady state is a continuous variable gaussian state completely characterised by $V(\infty)$.

\begin{figure*}[!t]
\includegraphics[width=1\textwidth]{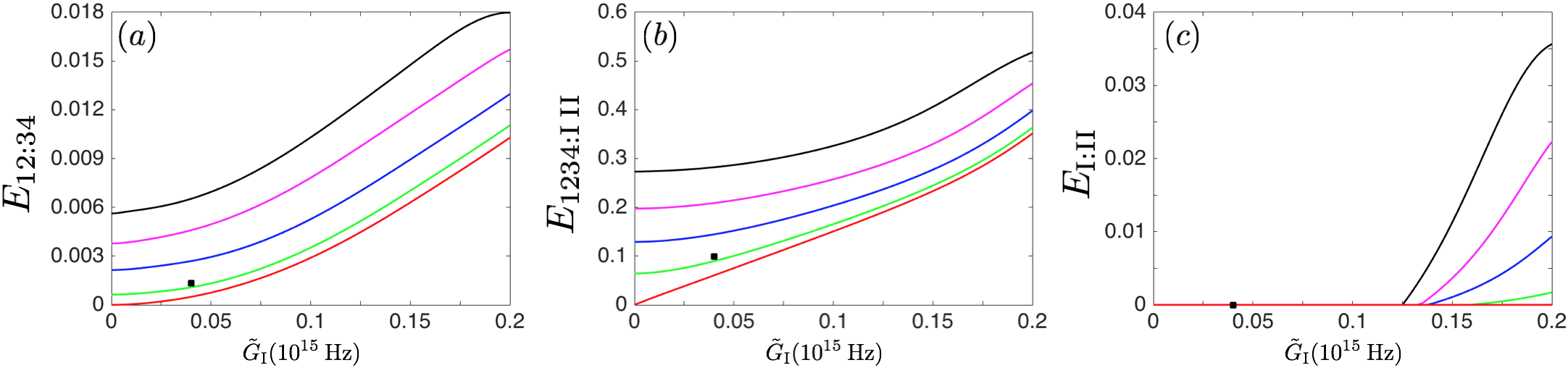} 
\caption{Steady-state entanglement (logarithmic negativity).
Entanglement between the cavity modes, panel (a), is always accompanied by considerable light-bacteria entanglement, panel (b), and for stronger couplings also by entanglement between the bacterial modes, panel (c).
In all cases, base coupling strengths are varied as $\tilde G_{{\scriptsize\mbox{I}}}=[0,0.2]\: 10^{15}\: \mbox{Hz}$ (horizontal axis) and $\tilde G_{{\scriptsize\mbox{II}}}=$ 0 (red lines), 0.05 (green lines), 0.1 (blue lines), 0.15 (magenta lines), and 0.2 (black lines) in $10^{15}\:\mbox{Hz}$. We have also indicated the experimentally realised coupling strength $\tilde G_{{\scriptsize\mbox{I}}}=3.9\times10^{13}\:\mbox{Hz}$ from Ref. \cite{bacteria-exp} and the corresponding $\tilde G_{{\scriptsize\mbox{II}}}=6\times10^{13}\:\mbox{Hz}$ as black dots.}
\label{FIG_ess}
\end{figure*}

\subsection{Results of calculations}
 
We now calculate the steady state entanglement using, wherever possible, parameters from the experiments of Ref.~\cite{bacteria-exp}.
We place the bacteria in a single-sided Fabry-Perot cavity of length $L=518$ nm (cf. Fig. \ref{FIG_ABC}). 
The refractive index due to aqueous bacterial solution embedded in the cavity is $n_r=\sqrt{\epsilon_r} \approx 1.33$, which gives the frequency of the $m^\text{th}$ cavity mode $\omega_m=m\pi c/n_rL\approx 1.37m\times 10^{15}$ Hz.
The reflectivities of the mirrors are engineered such that $R_2 = 100\%$ and $R_1 = 50\%$.
We assume the reflectivities are the same for all the optical modes, giving $\kappa_m \approx 7.5\times 10^{13}$ Hz through the finesse $\mathcal{F}=-2\pi/\ln{(R_1R_2)}=\pi c/2 \kappa_m n_r L$.
The decay rate of the excitons can be calculated as $\gamma_n=1/2\tau_n$ where $\tau_n=2h/\Gamma_n$ is the coherence time with $\Gamma_n$ being the full-width at half-maximum (FWHM) of the bacterial spectrum \cite{bajoni2012}. 
We approximate the spectrum in Fig. 1b of Ref. \cite{bacteria-exp} as a sum of two Lorentzian functions centred at $\Omega_{\scriptsize\mbox{I}}$ and $\Omega_{\scriptsize\mbox{II}}$ having FWHM of $(\Gamma_{\scriptsize \mbox{I}},\Gamma_{\scriptsize \mbox{II}})=(130,600)$ meV, giving $(\gamma_{\scriptsize\mbox{I}},\gamma_{\scriptsize\mbox{II}}) \approx (0.78,3.63)\times 10^{13}$ Hz respectively.
Note that the decay rate solely depends on the coherence time, i.e. we assume only homogenous broadening of the spectral lines.

All the spectral components of the driving laser are assumed to have the same power $P_m=50$mW and frequency $\Lambda_{m}=\omega_m$. 
By using the mode volume $V_m=2\pi L^3/m(1-R_1)$ \cite{modeV}, we can express the interaction strength as $G_{mn}= m\tilde G_n$, where we define $\tilde G_n \equiv \mu_n \sqrt{c(1-R_1)N/4\hbar n_r^3 \epsilon_0L^4}$. 
This quantity is a rate that characterises the base collective interaction strength of the cavity mode and the $n^{\scriptsize \mbox{th}}$ bacterial mode. 
Instead of fixing the value of $\tilde G_n$, we vary this quantity $\tilde G_n=[0,0.2]\: 10^{15}\: \mbox{Hz}$, which is within experimentally achievable regime (cf. Refs. \cite{bacteria-exp,bacteria-th}).

Logarithmic negativity is chosen as entanglement quantifier and the Methods section provides the details on how this quantity is calculated. 
We consider four cavity modes as the addition of higher modes shows negligible effects to the steady state entanglement.
In the steady state regime, we calculate entanglement between the cavity modes $E_{12:34}$, between the cavity modes and bacteria $E_{1234:{\scriptsize\mbox{I}}\:{\scriptsize\mbox{II}}}$, 
and between the bacterial modes $E_{{\scriptsize\mbox{I}}:{\scriptsize\mbox{II}}}$, cf. Fig. \ref{FIG_ess}. 
This steady state regime is reached in $\sim 100 \:\mbox{fs}$ (see Methods), which is faster than relaxation processes ($\sim$ ps) occuring within green sulphur bacteria \cite{chlorosomes}.
Our results show that the steady state entanglement $E_{12:34}$ is always accompanied by $E_{1234:{\scriptsize\mbox{I}}\:{\scriptsize\mbox{II}}}$, i.e. the bacteria are non-classically correlated with the cavity modes. 
This is in agreement with the general detection method of Ref.~\cite{revealing} as entanglement is a stronger type of quantum correlation than discord, i.e. nonzero $E_{1234:\scriptsize \mbox{I}\: \scriptsize \mbox{II}}$ implies nonzero cavity modes-bacteria discord $D_{1234|\scriptsize \mbox{I}\: \scriptsize \mbox{II}}$.
Our results also show that the entanglement dynamics of $E_{12:34}$ is dominated by modes $2$ and $3$ since other modes are further off resonance with the bacterial modes.
Moreover, there is entanglement generated within the bacteria.
This requires both $\tilde G_{{\scriptsize\mbox{I}}}$ and $\tilde G_{{\scriptsize\mbox{II}}}$ to be nonzero and relatively high.
We see that the bacteria can be strongly entangled with the cavity modes, much stronger than entanglement between the cavity modes. 
While the latter is in the order of $10^{-2} - 10^{-3}$, we note that entanglement in the range $10^{-2}$ has already been observed experimentally between mechanical motion and microwave cavity fields~\cite{evalue}.
We have also indicated, as black dotes in Fig. \ref{FIG_ess}, the coupling strengths $\tilde G_{{\scriptsize\mbox{I}}}=3.9\times10^{13}\:\mbox{Hz}$ from Ref. \cite{bacteria-exp} and the corresponding $\tilde G_{{\scriptsize\mbox{II}}}=6\times10^{13}\:\mbox{Hz}$, which is estimated as follows.
From the relation $\mu_n^2\propto \int f(\omega)d\omega /\omega_n$ \cite{houssier1970circular}, where $f$ is the extinction coefficient, one can obtain the ratio $\tilde G_{\scriptsize \mbox{II}}/\tilde G_{\scriptsize \mbox{I}}=\mu_{\scriptsize \mbox{II}}/\mu_{\scriptsize \mbox{I}}\approx 1.53$.

\section*{Discussion}

We point out that the covariance matrix $V(t)$, and hence the entanglement, does not depend on the power of the lasers.
This is a consequence of the dipole-dipole coupling and classical treatment of the driving field (see Methods).
Therefore, the system gets entangled also in the absence of the lasers.
There is no fundamental reason why this entanglement with vacuum could not be measured, but practically it is preferable to pump the cavity in order to improve the signal-to-noise ratio.
See also, e.g., Ref.~\cite{gyongyosi2017quantum} for efficient processing of post-measurement data.
Of course quantities other than entanglement may depend on driving power, for example the light intensity inside the cavity as shown in the Methods section.

This finding is quite different from results in optomechanical system where the covariance matrix depends on laser power \cite{paternostro,revealing}. 
The origin of this difference is the nature of the coupling. 
For example, in an optomechanical system consisting of a single cavity mode $\hat a$ and a mechanical mirror $\hat b$ the coupling is proportional to $\hat a^{\dagger}\hat a\hat x_b$, which is a third-order operator \cite{optmech1}. 
This results in the effective coupling strength being proportional to the classical cavity field intensity $\alpha$ after linearisation of the Langevin equations. 
This classical signal enters the covariance matrix via the effective coupling strength and introduces the dependence on the driving power. 

In order to justify the low atomic excitation limit we first note that the number of steady-state photons for the $m^{\scriptsize \mbox{th}}$ cavity mode without the presence of the bacteria is given by $E_m^2 / \kappa_m^2 \propto P_m$. 
When one considers the bacteria in the cavity having the base interaction strength $\tilde G_n$ and a decay rate $\gamma_n$ in the same order as the cavity decay rate, the number of excitation of the bacterial modes would also be in the order of $E_m^2 / \kappa_m^2 $, which in our case is $10^{3}$. 
With $\sim10^8$ actively coupled dipoles in the cavity \cite{bacteria-exp}, this gives $\sim10^{-3}\%$ excitation, which justifies the low-excitation approximation.
We also plotted the evolution of excitation numbers of the bacterial modes (together with the number of photons in different cavity modes) within our model, see Methods. 
It shows that excitation numbers are oscillating in the ``steady state". 
The oscillations are caused by the combination of interactions between the light and bacteria (Rabi-like oscillations) and the time-dependent driving laser.
Setting the interactions $G_{mn}=0$ or the driving off ($P_m=0$) indeed produces constant steady-state value.
We observe that the excitation number of the bacterial system is always bellow $2000$, which is in agreement with the statement above.

We also performed similar calculations in which we neglected the counter rotating terms in Eq. (\ref{EQ_H}), the model known as Tavis-Cummings. 
This resulted in no entanglement generated in the steady state and can be intuitively understood as follows.
Since the steady-state covariance matrix does not depend on the initial state and on the power of the driving lasers, we might start with all atoms in the ground state, vacuum for the light modes, and no driving.
Under such circumstances there is no interaction between bacterial modes and light modes as every term in the interaction Hamiltonian contains an annihilation operator.
In physical terms, since we begin with the lowest energy state and the interaction Hamiltonian preserves energy, the ground state will be the state of affairs at any time.
Therefore, nonzero entanglement observed in experiments will provide evidence of the counter rotating terms in the coupling.

\section*{Methods}\setcounter{subsection}{0}

\subsection{Evolution of quadratures}\label{APP_quadt}

The Langevin equations for the quadratures can be written in a simple matrix equation $\dot u(t)=Ku(t)+l(t)$, with the vector $u=( \hat x_{1}, \hat y_{1},\cdots, \hat x_{M}, \hat y_{M}, \hat x_{\scriptsize\mbox{I}}, \hat y_{\scriptsize\mbox{I}},\hat x_{\scriptsize\mbox{II}}, \hat y_{\scriptsize\mbox{II}})^T$ and
\begin{equation}
K=\left( \begin{array}{cccccc} 
I_1&\bm{0}&\cdots& \bm{0} &L_{1\scriptsize\mbox{I}}&L_{1\scriptsize\mbox{II}}\\
\bm{0}&I_2 &\cdots&\bm{0}&L_{2\scriptsize\mbox{I}}&L_{2\scriptsize\mbox{II}}\\
\vdots&\vdots&\ddots&\vdots&\vdots&\vdots\\
\bm{0}&\bm{0}&\cdots&I_{M}&L_{M\scriptsize\mbox{I}}&L_{M\scriptsize\mbox{II}}\\
L_{1\scriptsize\mbox{I}}&L_{2\scriptsize\mbox{I}}&\cdots&L_{M\scriptsize\mbox{I}}&I_{\scriptsize\mbox{I}}&\bm{0}\\
L_{1\scriptsize\mbox{II}}&L_{2\scriptsize\mbox{II}}&\cdots&L_{M\scriptsize\mbox{II}}&\bm{0}&I_{\scriptsize\mbox{II}}\\ 
\end{array}\right),
\end{equation}
where the components are $2\times2$ matrices given by
\begin{equation}
I_m=\left( \begin{array}{cc}  
-\kappa_m &\omega_m\\
-\omega_m&-\kappa_m\\
\end{array}\right), \: 
L_{mn}=\left( \begin{array}{cc}  
0 &0\\
-2G_{mn}&0\\
\end{array}\right), \nonumber
\end{equation}
\begin{equation} 
I_{n}=\left( \begin{array}{cc}  
-\gamma_n &\Omega_n\\
-\Omega_n&-\gamma_n\\
\end{array}\right),
\end{equation}
and $\bm{0}$ is a $2\times2$ zero matrix.
Note that we have used the index $m=1,2,\cdots,M$ for the cavity modes and $n=\scriptsize\mbox{I},\scriptsize\mbox{II}$ for the bacterial modes. 
We split the last term in the matrix equation into two parts, representing the noise and pumping respectively, i.e. $l(t)=\eta(t)+p(t)$ where 
\begin{equation}
\frac{\eta(t)}{\sqrt{2}}=\left( \begin{array}{c} 
\sqrt{\kappa_1}\: \hat X_1(t)\\ 
\sqrt{\kappa_1}\: \hat Y_1(t)\\ 
\vdots\\
\sqrt{\kappa_{M}}\: \hat X_{M}(t)\\ 
\sqrt{\kappa_{M}}\: \hat Y_{M}(t)\\ 
\sqrt{\gamma_{\scriptsize\mbox{I}}}\: \hat X_{\scriptsize\mbox{I}}(t)\\ 
\sqrt{\gamma_{\scriptsize\mbox{I}}}\: \hat Y_{\scriptsize\mbox{I}}(t)\\
\sqrt{\gamma_{\scriptsize\mbox{II}}}\: \hat X_{\scriptsize\mbox{II}}(t)\\ 
\sqrt{\gamma_{\scriptsize\mbox{II}}}\: \hat Y_{\scriptsize\mbox{II}}(t)\\
\end{array}\right),
\frac{p(t)}{\sqrt{2}}=\left( \begin{array}{c} 
E_{1}\cos{\Lambda_{1}t}\\ 
-E_{1}\sin{\Lambda_{1}t}\\ 
\vdots \\
E_{M}\cos{\Lambda_{M}t}\\ 
-E_{M}\sin{\Lambda_{M}t}\\ 
0\\ 
0\\
0\\ 
0\\
\end{array}\right).
\end{equation}
We have also used quadratures for the noise terms, i.e. through $\hat F_m=(\hat X_m+i\hat Y_m)/\sqrt{2}$ and $\hat Q_n=(\hat X_n+i\hat Y_n)/\sqrt{2}$.

The solution to the Langevin equations is given by 
\begin{eqnarray}\label{AEQ_Lsol}
u(t)&=&W_+(t)u(0)+W_+(t)\int_0^t dt^{\prime}  W_-(t^{\prime})l(t^{\prime}),
\end{eqnarray}
where $W_{\pm}(t)=\exp{(\pm Kt)}$. 
This allows numerical calculation of expectation value of the quadratures as a function of time, i.e. $\langle u_i(t)\rangle$ is given by the $i^{\text{th}}$ element of 
\begin{equation}\label{EQ_ui}
W_+(t)\langle u(0)\rangle+W_+(t)\int_0^t dt^{\prime}  W_-(t^{\prime})p(t^{\prime}),
\end{equation}
which is obtained as follows.
Since every component of $p(t)$ is not an operator, we have $\langle p_k(t)\rangle=\mbox{tr}(p_k(t)\rho)=p_k(t)$. 
Also, we have used the fact that the noises have zero mean, i.e. $\langle \eta_k(t)\rangle=0$.

\begin{figure*}[!t]
\includegraphics[width=0.95\textwidth]{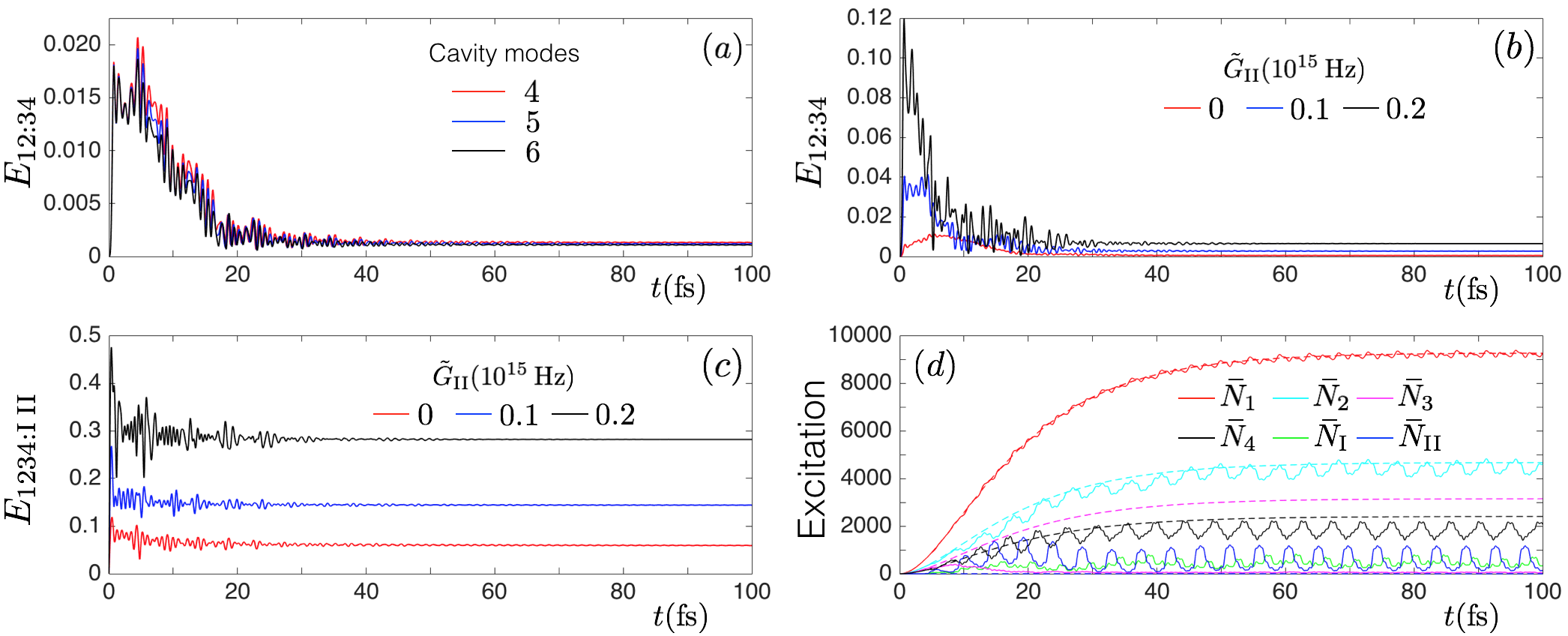}
\caption{Exemplary dynamics.
(a) Bipartite entanglement between the cavity modes $E_{12:34}$ taking into account up to $4$, $5$, and $6$ cavity modes, showing that higher modes do not contribute to steady state entanglement. 
(b) Entanglement in the partition $12:34$ with varying interaction strengths. 
(c) Entanglement between the cavity modes and bacteria, showing faster growth and much higher steady state values than entanglement between the cavity modes.
(d) Evolution of photon number of the cavity modes $\bar N_1$, $\bar N_2$, $\bar N_3$, $\bar N_4$ and excitation of the bacteria $\bar N_{\scriptsize\mbox{I}},\bar N_{\scriptsize\mbox{II}}$ (solid lines). Dashed lines represent the evolution when the interactions between the bacteria and light are absent ($G_{mn}=0$).
$\tilde G_{\scriptsize\mbox{II}}$ has been fixed to be $6 \times10^{13}\:\mbox{Hz}$ for (a) and (d) while $\tilde G_{\scriptsize\mbox{I}}=3.9\times 10^{13}\:\mbox{Hz}$ for all graphs. 
We considered four cavity modes in (b), (c), and (d).
In all cases above, steady state entanglement is reached in $\sim 100$ fs.}
\label{FIG_evstime} 
\end{figure*}

\subsection{Covariance matrix}\label{APP_CM}

Covariance matrix of our system is defined as $V_{ij}(t)\equiv \langle \{ \Delta u_i(t),\Delta u_j(t)\}\rangle/2=\langle u_i(t)u_j(t)+u_j(t)u_i(t)\rangle/2-\langle u_i(t)\rangle \langle u_j(t)\rangle$ where we have used $\Delta u_i(t)=u_i(t)-\langle u_i(t)\rangle$. 
This means that $p(t)$ does not contribute to $\Delta u_i(t)$ (and hence the covariance matrix) since $\langle p_k(t)\rangle=p_k(t)$. 
We can then construct the covariance matrix at time $t$ from Eq. (\ref{AEQ_Lsol}) without considering $p(t)$ as follows
\begin{eqnarray}
V_{ij}(t)&=&\langle u_i(t)u_j(t)+u_j(t)u_i(t)\rangle/2-\langle u_i(t)\rangle \langle u_j(t)\rangle \nonumber \\
\label{EQ_l3}
V(t)&=&W_+(t)V(0)W_+^T(t) \nonumber \\
&&+W_+(t)\int_0^t dt^{\prime} W_-(t^{\prime})DW_-^T(t^{\prime}) \:W_+^T(t) ,
\end{eqnarray}
where $D=\mbox{Diag}[\kappa_1,\kappa_1,\cdots,\kappa_M,\kappa_M,\gamma_{\scriptsize\mbox{I}},\gamma_{\scriptsize\mbox{I}},\gamma_{\scriptsize\mbox{II}},\gamma_{\scriptsize\mbox{II}}]$ and we have assumed that the initial quadratures are not correlated with the noise quadratures such that the mean of the cross terms are zero. 
A more explicit solution of the covariance matrix, after integration in Eq. (\ref{EQ_l3}), is given by 
\begin{eqnarray}\label{EQ_Ct}
KV(t)+V(t)K^T&=&-D+KW_+(t)V(0)W_+^T(t) \nonumber \\
&&+W_+(t)V(0)W_+^T(t)K^T \nonumber \\
&&+W_+(t)DW_+^T(t),
\end{eqnarray}
which is linear and can be solved numerically. 

The steady state is guaranteed when all real parts of the eigenvalues of $K$ are negative, i.e. $W_+(\infty)=\bm{0}$.
In this case the covariance matrix satisfies Eq. (\ref{EQ_Css}).

\subsection{Entanglement from covariance matrix}
\label{APP_ENT}

The covariance matrix $V$ describing our system can be written in block form 
\begin{equation}\label{EQ_COV}
V=\left( \begin{array}{cccc} 
B_{11}&B_{12}&\cdots&B_{1Z}\\
B_{12}^T&B_{22}&\cdots&B_{2Z}\\ 
\vdots&\vdots&\ddots&\vdots \\ 
B_{1Z}^T&B_{2Z}^T&\cdots&B_{ZZ}\\
\end{array}\right),
\end{equation}
where $Z$ is the total number of modes, which is $M+2$ in our case. 
The block component, here denoted as $B_{jk}$, is a $2\times 2$ matrix describing local mode correlation when $j=k$ and intermodal correlation when $j\ne k$.
A $Z$-mode covariance matrix has symplectic eigenvalues $\{\nu_k\}_{k=1}^{Z}$ that can be computed from the spectrum of matrix $|i\Omega_{Z} V|$ \cite{weedbrook2012gaussian} where 
\begin{equation}
 \Omega_{Z}=\bigoplus^{Z}_{k=1} \left( \begin{array}{cc} 0&1\\ -1 &0\end{array}\right).
 \end{equation}
For a physical covariance matrix $2 \nu_k\ge 1$. 

Entanglement is calculated as follows. For example, the calculation in the partition $12:34$ only requires the covariance matrix of modes $1,2,3$, and $4$:
\begin{equation}
V=\left( \begin{array}{cccc} 
B_{11}&B_{12}&B_{13}&B_{14}\\
B_{12}^T&B_{22}&B_{23}&B_{24}\\ 
B_{13}^T&B_{23}^T&B_{33}&B_{34} \\ 
B_{14}^T&B_{24}^T&B_{34}^T&B_{44}\\
\end{array}\right),
\end{equation}
that can be obtained from Eq. (\ref{EQ_COV}).
If the covariance matrix $\tilde V$, after partial transposition with respect to mode $3$ and $4$ (this is equivalent to flipping the sign of the operator $\hat y_3$ and $\hat y_4$ in $V$) is not physical, then our system is entangled. 
This unphysical $\tilde V$ is shown by its minimum symplectic eigenvalue $\tilde \nu_{\mbox{min}}<1/2$.
Entanglement is then quantified by logarithmic negativity as follows $E_{12:34}=\mbox{max}[0,-\ln{(2\tilde \nu_{\mbox{min}})}]$ \cite{negativity, adesso2004extremal}. 
Note that the separability condition, when $\tilde \nu_{\mbox{min}}\ge1/2$, is sufficient and necessary when one considers bipartitions with one mode on one side \cite{werner2001bound}, e.g. partition between bacterial modes $\mbox{I}:\mbox{II}$. 

\subsection{Dynamics of entanglement and excitation numbers}
\label{APP_ED}

Let us consider as initial the time right before the bacteria are inserted into the cavity.
Then all the cavity modes and the bacteria are completely uncorrelated and do not interact.
The dynamics is then started by placing the bacteria in the cavity.
In what follows, as an example of the dynamics we start with vacuum state for the cavity modes and ground state for the bacteria. 
The initial state of the bacteria is justified by the fact that $\hbar \Omega_n \gg k_BT$, even at room temperature. 

Fig. \ref{FIG_evstime} (a)-(c) show the resulting entanglement dynamics.
Panel (a) displays existence of steady-state entanglement between cavity modes $1,2$ and $3,4$, which is not altered heavily if the calculations take into account five and six cavity modes in total. 
Therefore, we consider $4$ cavity modes in all other calculations.
In recent experiments, the rate $\tilde G_{\scriptsize \mbox{I}}$ was shown to be $3.9\times 10^{13}\:\mbox{Hz}$ \cite{bacteria-exp} and the corresponding $\tilde G_{\scriptsize \mbox{II}}=6\times 10^{13}\:\mbox{Hz}$.
In our calculations we vary this rate as in panels (b) and (c) (also see Fig. \ref{FIG_ess}).
As expected the higher the rate the more entanglement gets generated.
It is also apparent that entanglement between the cavity modes and bacteria $E_{1234:\scriptsize \mbox{I}\: \scriptsize \mbox{II}}$ grows faster than entanglement between the cavity modes.
More precisely, nonzero $E_{12:34}$ implies nonzero $E_{1234:\scriptsize \mbox{I}\: \scriptsize \mbox{II}}$.

The excitation number of the cavity modes and bacteria as a function of time can be calculated from $\langle u_i(t)\rangle$ and $V_{ii}(t)$. For example, the mean excitation number for the first cavity mode is given by
\begin{eqnarray}
\bar N_{1}(t)=\langle \hat a_1^{\dagger}(t)\hat a_1(t)\rangle &=&\frac12 ( V_{11}(t)+V_{22}(t) \nonumber \\
&&+\langle u_1(t)\rangle^2+\langle u_2(t)\rangle^2-1). \nonumber \\
&&
\end{eqnarray}

We present the evolution of photon number of the cavity modes and excitation of the bacterial modes in Fig. \ref{FIG_evstime} (d). Note that photon number of the $3^\text{rd}$ cavity mode (solid magenta line) is showing oscillations well bellow its ``off-interaction" value (dashed magenta line).
This is because $\omega_3$ is almost in resonance with the frequency of the atomic transition $\Omega_{\scriptsize \mbox{II}}$.

%\section*{Data availability} 
%This work has no data.

\section*{Acknowledgments} 

We thank David Coles for consultations and experimental data.
We also thank anonymous referees for their comments and suggestions.
This work is supported by the National Research Foundation (Singapore) and Singapore Ministry of Education Academic Research Fund Tier 2 Project No. MOE2015-T2-2-034. 
CM is supported by the Templeton World Charity Foundation and the Eutopia Foundation. MP is supported by the SFI-DfE Investigator Programme (Grant 15/IA/2864) and a Royal Society Newton Mobility Grant (NI160057).

%\section*{Competing interests}
%The authors have no financial or non-financial conflicts of interest.

%\section*{Author contributions}
%All authors researched, collated, and wrote this paper.


\begin{thebibliography}{99}

\bibitem{SCH} Schr\"{o}dinger, E.
\newblock {\em What is life?}
\newblock (University Press, Cambridge, 1943).

\bibitem{BOH} Bohr, N.
\newblock Light and Life.
\newblock {\em Nature} {\bf 131}, 457 (1933).

\bibitem{WIG} Wigner, E.
\newblock {\em The Probability of a Self-Reproducing Unit. Symmetries} \& {\em Reflections}
\newblock (Indiana University Press, Bloomington, 1967).

\bibitem{arndt} Gerlich, S. \emph{et al}.
\newblock Quantum interference of large organic molecules.
\newblock {\em Nat. Commun.} {\bf 2}, 263 (2011).

\bibitem{discord}  Henderson, L. \& Vedral, V.
\newblock Classical, quantum and total correlations.
\newblock {\em J. Phys. A} {\bf 34}, 6899 (2001).

\bibitem{discord2} Ollivier, H. \& Zurek, W. H.
\newblock Quantum discord: a measure of the quantumness of correlations.
\newblock {\em Phys. Rev. Lett.} {\bf 88}, 017901 (2001).

\bibitem{discord3} Celeri, L. C., Maziero, J. \& Serra, R. M. 
\newblock Theoretical and experimental aspects of quantum discord and related measures.
\newblock{\em Int. J. Quant. Inf.} {\bf 9}, 1837 (2011).

\bibitem{discord4} Modi, K., Brodutch, A., Cable, H., Paterek, T. \& Vedral, V.
\newblock The classical-quantum boundary for correlations: discord and related measures.
\newblock{\em Rev. Mod. Phys.} {\bf 84}, 1655 (2012).

\bibitem{discord5} Adesso, G., Bromley, T. R. \& Cianciarusso, M.
\newblock Measures and applications of quantum correlations.
\newblock{\em J. Phys. A} {\bf 49}, 473001 (2016).

\bibitem{coherence-distillation} Chitambar, E. \emph{et al}.
\newblock Assisted distillation of quantum coherence.
\newblock{\em Phys. Rev. Lett.} {\bf 116}, 070402 (2016).

\bibitem{bacteria-th} Marletto, C., Coles, D., Farrow, T. \& Vedral, V.
\newblock Entanglement between living bacteria and quantized light witnessed by Rabi splitting.
\newblock {\em J. Phys. Commun.} {\bf 2}, 101001 (2018)

\bibitem{bacteria-exp} Coles, D. \emph{et al}.
\newblock A Nanophotonic Structure Containing Living Photosynthetic Bacteria.
\newblock {\em Small} {\bf 13}, 1701777 (2017).

\bibitem{bio1} Amerongen, H., Valkunas, L. \& Grondelle, R. 
\newblock {\em Photosynthetic excitons}
\newblock (World Scientific, Singapore, 2000).

\bibitem{bio2} Lee, H., Cheng, Y. C. \& Fleming, G. R.
\newblock Coherence dynamics in photosynthesis: Protein protection of excitonic coherence.
\newblock {\em Science} {\bf 316}, 1462 (2007).

\bibitem{bio3} Engel, G. S. \emph{et al}.
\newblock Evidence for wavelike energy transfer through quantum coherence in photosynthetic systems.
\newblock {\em Nature} {\bf 446}, 782 (2007).

\bibitem{bio4} Sarovar, M., Ishizaki, A., Fleming, G. R. \& Whaley, K. B.
\newblock Quantum entanglement in photosynthetic light-harvesting complexes.
\newblock {\em Nat. Phys.} {\bf 6}, 462 (2010).

\bibitem{bio5} Collini, E. \emph{et al}.
\newblock Coherently wired light-harvesting in photosynthetic marine algae at ambient temperature.
\newblock {\em Nature} {\bf 463}, 644 (2010).

\bibitem{bio6} Panitchayangkoon, G. \emph{et al}.
\newblock Long-lived quantum coherence in photosynthetic complexes at physiological temperature.
\newblock {\em Proc. Natl Acad. Sci. USA} {\bf 107}, 12766 (2010).

\bibitem{bio7} Wilde, M. M., McCracken, J. M. \& Mizel, A.
\newblock Could light harvesting complexes exhibit non-classical effects at room temperature?
\newblock {\em Proc. R. Soc. A} {\bf 446}, 1347 (2010).

\bibitem{bio8} Li, C. M., Lambert, N., Chen, Y. N., Chen, G. Y. \& Nori, F.
\newblock Witnessing Quantum Coherence: from solid-state to biological systems.
\newblock {\em Sci. Rep.} {\bf 2}, 885 (2012).

\bibitem{bio9} Hildner, R., Brinks, D., Nieder, J. B., Cogdell, R. J. \& Hulst, N. F.
\newblock Quantum coherent energy transfer over varying pathways in single light-harvesting complexes.
\newblock {\em Science} {\bf 340}, 1448 (2013).

\bibitem{bio10} Lambert, N. \emph{et al}.
\newblock Quantum biology.
\newblock {\em Nat. Phys.} {\bf 9}, 10 (2013).

\bibitem{bio11} Scholes, G. D. \emph{et al}.
\newblock Using coherence to enhance function in chemical and biophysical systems.
\newblock {\em Nature.} {\bf 543}, 647 (2017).

\bibitem{rabienergy} Fox, M.
\newblock {\em Quantum Optics-An Introduction. Oxford Master Series}
\newblock (Oxford University Press, New York, 2006).

\bibitem{ZHU} Zhu, Y. \emph{et al}.
\newblock Vacuum Rabi splitting as a feature of linear-dispersion theory: Analysis and experimental observations.
\newblock {\em Phys. Rev. Lett.} {\bf 64}, 21 (1990).

\bibitem{revealing} Krisnanda, T., Zuppardo, M., Paternostro, M. \& Paterek, T.
\newblock Revealing Nonclassicality of Inaccessible Objects.
\newblock {\em Phys. Rev. Lett.} {\bf 119}, 120402 (2017).

\bibitem{gyongyosi2018survey} Gyongyosi, L., Imre, S. \& Nguyen, H. V.
\newblock A survey on quantum channel capacities.
\newblock {\em IEEE Commun. Surv. Tut.} {\bf 20}, 1149 (2018).

\bibitem{imre2012advanced} Imre, S. \& Gyongyosi, L.
\newblock{\em Advanced quantum communications: an engineering approach}
\newblock (John Wiley \& Sons, New Jersey, 2012).

\bibitem{blankenship1995antenna} Blankenship, R. E., Olson, J. M. \& Miller, M.
\newblock {\em Anoxygenic photosynthetic bacteria}
\newblock (Springer, Dordrecht, 1995).

\bibitem{chlorosomes} Coles, D. \emph{et al}.
\newblock Strong coupling between chlorosomes of photosynthetic bacteria and a confined optical cavity mode.
\newblock {\em Nat. Commun.} {\bf 5}, 5561 (2014).

\bibitem{HP} Holstein, T. \& Primakoff, H.
\newblock Field dependence of the intrinsic domain magnetization of a ferromagnet.
\newblock {\em Phys. Rev.} {\bf 58}, 1098 (1940).

\bibitem{optoatoms} Genes, C., Vitali, D. \& Tombesi, P.
\newblock Emergence of atom-light-mirror entanglement inside an optical cavity.
\newblock {\em Phys. Rev. A} {\bf 77}, 050307 (2008).

\bibitem{bai2016robust} Bai, C. H., Wang, D. Y., Wang, H. F., Zhu, A. D. \& Zhang, S.
\newblock Robust entanglement between a movable mirror and atomic ensemble and entanglement transfer in coupled optomechanical system.
\newblock{\em Sci. Rep.} {\bf 6}, 33404 (2016).

\bibitem{noise} Gardiner, C. W. \& Zoller, P.
\newblock {\em Quantum noise}
\newblock (Springer Science \& Business Media, Berlin, 2004).

\bibitem{noise2} Walls, D. F. \& Milburn, G. J.
\newblock{\em Quantum optics}
\newblock (Springer Science \& Business Media, Berlin, 2007).

\bibitem{bajoni2012} Bajoni, D.
\newblock Polariton lasers. Hybrid light-matter lasers without inversion.
\newblock {\em J. Phys. D: Appl. Phys.} {\bf 45}, 313001 (2012).

%\bibitem{bacmodel} We approximate the spectrum in Fig. 1b of Ref. \cite{bacteria-exp} as a sum of two Lorentzian functions centred at $750$ and $460$ nm having FWHM of $130$ and $600$ meV respectively. Also, the relative base interaction strength $\tilde G_{\scriptsize \mbox{II}}/\tilde G_{\scriptsize \mbox{I}}=\mu_{\scriptsize \mbox{II}}/\mu_{\scriptsize \mbox{I}}\approx 1.53$ is obtained using the relation $\mu_n^2\omega_n\propto \int f(\omega)d\omega$ \cite{houssier1970circular}, where $f$ is the extinction coefficient.

\bibitem{modeV} Ujihara, K.
\newblock Spontaneous emission and the concept of effective area in a very short optical cavity with plane-parallel dielectric mirrors.
\newblock {\em Jpn. J. Appl. Phys.} {\bf 30}, L901 (1991).
\newblock Note that cavities with different geometries will have different mode volume formulae.

\bibitem{evalue} Palomaki, T. A., Teufel, J. D., Simmonds,  R. W. \& Lehnert, K. W.
\newblock Entangling mechanical motion with microwave fields.
\newblock {\em Science} {\bf 342}, 710 (2013).

\bibitem{houssier1970circular} Houssier, C. \& Sauer, K.
\newblock Circular dichroism and magnetic circular dichroism of the chlorophyll and protochlorophyll pigments.
\newblock {\em J. Am. Chem. Soc.} {\bf 92}, 779 (1970).

\bibitem{gyongyosi2017quantum} Gyongyosi, L.
\newblock Quantum imaging of high-dimensional Hilbert spaces with Radon transform.
\newblock {\em Int. J. Circ. Theor. App.} {\bf 45}, 1029 (2017).

\bibitem{paternostro} Paternostro, M. \emph{et al}.
\newblock Creating and probing multipartite macroscopic entanglement with light.
\newblock {\em Phys. Rev. Lett.} {\bf 99}, 250401 (2007).

\bibitem{optmech1} Vitali, D. \emph{et al}.
\newblock Optomechanical entanglement between a movable mirror and a cavity field.
\newblock {\em Phys. Rev. Lett.} {\bf 98}, 030405 (2007).

\bibitem{weedbrook2012gaussian} Weedbrook, C. \emph{et al}.
\newblock Gaussian quantum information.
\newblock {\em Rev. Mod. Phys.} {\bf 84}, 621 (2012).

\bibitem{negativity} Vidal, G. \& Werner, R. F.
\newblock Computable measure of entanglement.
\newblock {\em Phys. Rev. A} {\bf 65}, 032314 (2002).

\bibitem{adesso2004extremal} Adesso,  G., Serafini,  A. \& Illuminati, F.
\newblock Extremal entanglement and mixedness in continuous variable systems.
\newblock {\em Phys. Rev. A} {\bf 70}, 022318 (2004).

\bibitem{werner2001bound} Werner, R.~F. \& Wolf, M. M.
\newblock Bound entangled gaussian states.
\newblock {\em Phys. Rev. Lett.} {\bf 86}, 3658 (2001).

\end{thebibliography}
\end{document}